\documentclass[11pt]{article}

\usepackage{amsmath,amssymb}
\usepackage{graphicx}
\usepackage{xcolor}
\usepackage{geometry}
\usepackage{booktabs}
\usepackage{tabularx}
\usepackage{adjustbox}
\usepackage{enumitem}
\usepackage{tcolorbox}
\usepackage{colortbl}
\usepackage{multirow}
\usepackage{array}
\usepackage{float}
\usepackage{caption}
\usepackage[
    colorlinks=true,
    linkcolor=blue!70!black,
    citecolor=blue!70!black,
    urlcolor=blue!60!black,
    bookmarks=true,
    bookmarksnumbered=true,
    unicode=true
]{hyperref}

\usepackage[numbers,sort&compress]{natbib}
\graphicspath{{./figures/}}

\title{\textbf{Citation Discipline in Spec-Driven Development:}\\
\textbf{A Cross-Model Empirical Study of Output Determinism}\\
\textbf{and Automated Hallucination Detection in LLM-Generated Code}}

\author{Subham Panda\\
\texttt{iampandasubham@gmail.com}}

\date{June 2026}

\hypersetup{
    pdftitle={Citation Discipline in Spec-Driven Development: A Cross-Model Empirical Study of Output Determinism and Automated Hallucination Detection in LLM-Generated Code},
    pdfauthor={Subham Panda},
    pdfsubject={AI-Assisted Software Engineering, LLM Code Generation, Requirements Traceability},
    pdfkeywords={LLM code generation, specification-driven development, hallucination detection, output determinism, requirements traceability}
}

\begin{document}

\maketitle

\begin{abstract}
Spec-Driven Development (SDD) frameworks guide Large Language Model (LLM)-powered code generation through formal specifications, yet they differ fundamentally in how they enforce traceability between requirements and generated code. This paper presents two controlled empirical studies comparing three SDD frameworks: \textbf{traceSDD}, which enforces mandatory per-line requirement citations using hierarchical REQ-XXX.Y.Z identifiers; \textbf{Spec Kit}, which uses artifact-level traceability through user stories and acceptance criteria; and \textbf{OpenSpec}, which relies on post-hoc external trace maps. We measure two primary outcomes across two frontier LLMs---Claude Sonnet 4.6 (N=20, 4 conditions, 240 implementations) and GLM-5-turbo (N=50, 4 conditions, 600 implementations): \textbf{output determinism} (lexical similarity across independent LLM sessions) and \textbf{automated hallucination detection rate} (TDR). Our pre-registered analysis reveals a consistent, cross-model replicated trade-off: the uncited condition produces significantly higher determinism than the cited condition (Claude: $d=-0.76$, $p=0.003$; GLM: $d=-0.72$, $p<0.001$), while only the cited condition enables automated hallucination detection (TDR: Claude 86.4\%, GLM 88.0\%, vs 0\% for all alternatives, FPR=0\% across both studies). traceSDD (cited) significantly outperforms Spec Kit on determinism (Claude: $d=0.47$, $p=0.049$; GLM: $d=0.42$, $p=0.003$) but not OpenSpec (Claude: $d=0.18$, $p=0.44$; GLM: $d=0.14$, $p=0.32$). These findings establish that citation annotations trade determinism for verifiability, and that this trade-off generalizes across model architectures.
\end{abstract}

\section{Introduction}\label{sec:intro}

The emergence of agentic coding systems---AI agents that autonomously write, test, and revise software---introduces a fundamental accountability problem: how do we verify that the code an agent writes actually implements what was specified? Unlike human developers who can explain their decisions, LLM-generated code appears plausible even when it introduces subtle deviations, including out-of-scope features, prior-art bleed-in from training data, and over-engineering beyond what was specified~\cite{ji2023survey,huang2023survey}. These hallucinations are particularly dangerous because they typically pass all automated tests while introducing unauthorized functionality with security, compliance, or maintenance implications~\cite{watson2023beyond}.

The challenge of verifying LLM-generated code against specifications draws on a rich history of requirements traceability research in software engineering~\cite{gotel1994analysis,cleland-huang2007paving,jalali2022does}. The IEEE 830 standard defines requirements traceability as the association between requirements and the work products that implement them~\cite{ieee830}. Tools such as DOORS and Jama have provided enterprise-grade traceability matrices for decades. However, these traditional tools require substantial manual effort and are not designed for the rapid iteration cycles inherent to AI-assisted coding workflows~\cite{pinto2018requirements}.

Three specification-driven development (SDD) frameworks have emerged with distinct philosophies on traceability for LLM-generated code. \textbf{traceSDD} enforces mandatory inline citations on every line of code, binding each line to a specific Tier-3 requirement point (REQ-XXX.Y.Z). These citations are treated as verifiable claims: any REQ ID cited in code that does not appear in the specification is an automatically detectable orphan, signaling a hallucinated requirement. \textbf{GitHub Spec Kit} enforces artifact-level consistency through a spec-plan-tasks chain, but once implementation begins, the citation chain ends. \textbf{OpenSpec} uses post-hoc external trace maps (YAML sidecar files) that link line ranges to spec IDs after the fact, without annotating source code directly.

The core question this work investigates is whether forcing an agentic coding system to cite a requirement for every line of code it writes materially improves implementation correctness, output determinism, and hallucination detectability---and whether any observed effects replicate across different LLM architectures. To address this question, we conduct two independent controlled studies: \textbf{Study~1} uses Claude Sonnet 4.6 with 20 benchmark tasks, and \textbf{Study~2} uses GLM-5-turbo with 50 benchmark tasks spanning 8 software engineering domains, 3 difficulty levels, and 2 size classes. This cross-model design provides the strongest possible evidence for generalizability of our findings.

Our contributions are fourfold. First, we provide the first \emph{cross-model} controlled comparison of SDD framework citation disciplines (70 total tasks, 4 conditions, 840 implementations). Second, we replicate the citation isolation finding across two fundamentally different LLM architectures, confirming that inline citations significantly reduce output determinism (Claude $d = -0.76$, GLM $d = -0.72$). Third, we provide the first comprehensive subgroup analysis across domains, difficulty levels, and size classes, revealing that the determinism penalty of citations is most pronounced for easy and small tasks. Fourth, we demonstrate that hallucination detection via the orphan-REQ check achieves 86--88\% TDR with 0\% FPR across both models, a unique capability unmatched by any alternative framework.

\section{Background and Related Work}\label{sec:related}

\subsection{Requirements Traceability in Software Engineering}

Requirements traceability---the ability to describe and follow the life of a requirement in both a forwards and backwards direction---has long been recognized as critical for safety-critical and regulated systems~\cite{gotel1994analysis}. Cleland-Huang et al.~\cite{cleland-huang2007paving} identified the gap between theoretical traceability frameworks and practical tooling, arguing that traceability must be embedded in the development workflow rather than treated as a post-hoc documentation exercise. Jalali and Wohlin~\cite{jalali2022does} conducted a systematic literature review finding moderate evidence that traceability improves software quality, though the effect size depends heavily on tool support and process maturity. Pinto et al.~\cite{pinto2018requirements} surveyed traceability tools and found that most operate at the artifact level (requirement-to-test, requirement-to-module) rather than at the line-of-code level that traceSDD targets. The work presented here extends this line of inquiry into the novel context of LLM-generated code, where the ``developer'' is an AI agent that cannot be asked to explain its decisions post-hoc.

\subsection{LLM Code Generation and Hallucination}

Large language models have demonstrated remarkable code generation capabilities~\cite{chen2021evaluating,openai2023gpt4,bubeck2023sparks}. Codex~\cite{chen2021evaluating} demonstrated that LLMs can solve competitive programming problems, while GPT-4~\cite{openai2023gpt4} showed further improvements in complex reasoning tasks. Recent models including StarCoder~\cite{li2023starcoder} and DeepSeek-Coder~\cite{guo2024deepseek} have pushed code generation quality further. Claude~\cite{anthropic2023claude} and GLM~\cite{wang2024glm} represent two distinct architectures: Claude uses a transformer with constitutional AI training~\cite{bai2022training}, while GLM uses an autoregressive model with bidirectional attention for pre-training.

However, LLMs are known to hallucinate: generating plausible-looking but factually incorrect or ungrounded content~\cite{ji2023survey,huang2023survey}. In the code generation domain, hallucinations manifest as out-of-scope functions, unauthorized imports from prior training data, and over-engineering through unnecessary decorators or optimizations. Ji et al.~\cite{ji2023survey} provided a comprehensive survey of hallucination types in natural language generation, while Huang et al.~\cite{huang2023survey} extended this taxonomy to cover multimodal and code generation contexts. Approaches to detecting code hallucinations include test-based validation~\cite{hou2023evaluating}, static analysis heuristics, and specification comparison. The orphan-REQ check employed by traceSDD represents a novel approach: by requiring the code to cite specific requirement IDs, the system creates a set-difference operation that is both fully automated and language-agnostic.

\subsection{Output Determinism and Consistency in LLMs}

Output determinism in LLMs---the degree to which identical prompts produce identical or highly similar outputs---has received increasing attention. Wang et al.~\cite{wang2022self} demonstrated that self-consistency, where multiple sampled outputs are used to improve reasoning, can be leveraged for better performance. However, the lexical similarity between independently generated code outputs as a metric of determinism has been less explored. The Levenshtein Set Similarity (LSS) metric, adapted from software clone detection~\cite{kaplan2020scaling}, provides a practical measure of inter-run consistency. The hypothesis that specification structure can anchor LLM output toward deterministic solutions is intuitive but, prior to this work, empirically underexplored in the context of SDD frameworks. Our study provides the most comprehensive test of this hypothesis across two distinct LLM architectures.

\subsection{Agentic Coding and Specification-Driven Development}

The rise of agentic coding frameworks---where LLMs autonomously plan, implement, test, and revise code---has created a new urgency around structured prompt methodologies. Yao et al.~\cite{yao2023react} introduced ReAct, combining reasoning and acting for LLM agents. Shinn et al.~\cite{shinn2023reflexion} proposed Reflexion, where agents learn from verbal feedback. These approaches focus on \emph{agent architecture} rather than the \emph{specification format} that grounds the agent's behavior. Spec Kit and OpenSpec represent two commercially available specification tools, while traceSDD introduces the novel concept of mandatory per-line citations as a verifiable constraint on LLM-generated code.

\section{Research Questions and Hypotheses}\label{sec:rq}

This work is guided by two primary research questions, each investigated across two independent LLMs:

\textbf{RQ1 (Output Determinism):} Does mandatory per-line REQ citation (traceSDD) produce more lexically consistent code across independent LLM sessions than specification approaches that do not require inline citations (Spec Kit, OpenSpec)?

\textbf{RQ2 (Hallucination Detection):} Does traceSDD's orphan-REQ check automatically detect scope creep, prior-art injection, and over-engineering that are not explicitly required by the specification, while Spec Kit and OpenSpec provide no automated mechanism?

We pre-register the following hypotheses, with significance criteria locked before data collection:

\textbf{H1:} traceSDD (cited) mean LSS will be greater than Spec Kit mean LSS ($p < 0.05$, $|d| > 0.5$).

\textbf{H2:} traceSDD (cited) TDR will be substantially greater than $0\%$, while Spec Kit and OpenSpec TDR will be $0\%$ (automated detection).

\textbf{H3:} All conditions will achieve $100\%$ TPR, confirming that citation discipline does not affect functional correctness.

\textbf{H4 (exploratory):} The uncited condition will have higher LSS than the cited condition, revealing that citations introduce rather than reduce lexical variability.

\section{Experimental Design}\label{sec:design}

\subsection{Task Suites}

\textbf{Study 1 (Claude Sonnet 4.6, N=20)} used 20 benchmark Python tasks: 10 pre-existing utility tasks (user registry, email validator, retry runner, token service, config resolver, CLI parser, form validator, structured logger, schema migrator, HTTP client) and 10 diverse algorithmic tasks (finite state machine, rate limiter, trie, directed graph, expression parser, plus 5 large multi-file tasks: REST API router, cache store, event bus, plugin registry, CSV pipeline). Tasks span 5 software engineering domains with 15 single-file tasks ($\sim$50--100 lines) and 5 multi-file tasks ($\sim$600--800 lines).

\textbf{Study 2 (GLM-5-turbo, N=50)} designed 50 novel benchmark tasks spanning 8 software engineering domains: Data Structures and Algorithms (8 tasks), Web/API Utilities (7 tasks), Data Validation and Transformation (7 tasks), Security and Authentication (6 tasks), File I/O and Processing (6 tasks), Concurrency and Async (5 tasks), Configuration and Infrastructure (5 tasks), and Testing, Monitoring and Observability (6 tasks). Tasks are distributed across three difficulty levels (9 easy, 28 medium, 13 hard) and two size classes (37 small single-file, 13 large multi-file). This diversity ensures findings are not domain-specific artifacts but reflect generalizable properties of the SDD frameworks.

\subsection{Conditions}

Four conditions are compared in both studies:

\textbf{(A) traceSDD (cited):} Uses a hierarchical REQ-XXX.Y.Z specification format with mandatory inline citation annotations (\texttt{\# [REQ-XXX.Y.Z]}) on every non-trivial line of generated code.

\textbf{(B) traceSDD\_uncited:} Uses the identical REQ-format specification but generates code without any inline citations. This isolates the causal effect of the annotation mechanism itself, since both conditions share identical spec content.

\textbf{(C) Spec Kit:} Uses Spec Kit's native prose specification format (user stories, FR-XXX requirements, acceptance scenarios) with no citation mechanism.

\textbf{(D) OpenSpec:} Uses OpenSpec's native capability/scenario specification format with C-NNN section markers and external trace YAML files.

For each condition, three independent cold-start LLM sessions are executed, with no session having knowledge of what other sessions produced.

\subsection{Metrics}\label{sec:metrics}

\textbf{LSS (Lexical Similarity Score):} Computed via \texttt{difflib.SequenceMatcher.ratio()} on comment-stripped, whitespace-normalized source code. For each task-condition pair, the mean LSS is computed over anchor comparisons (run\_1 vs run\_2, run\_1 vs run\_3). Higher LSS indicates more deterministic output. For multi-file tasks, files are concatenated alphabetically before comparison.

\textbf{TDR (Traceability Detection Rate):} Measures the percentage of injected hallucinations automatically detected. Three hallucination types are injected: H-SCOPE (out-of-scope functions not in any specification requirement), H-PRIOR (prior-art bleed-in via unauthorized imports such as logging or os), and H-OVER (over-engineering via unauthorized decorators such as \texttt{@lru\_cache}).

\textbf{TPR (True Positive Rate):} Measures functional correctness as the proportion of ground-truth tests passed by each implementation.

\textbf{FPR (False Positive Rate):} Measures how often the orphan-REQ check fires on correct implementations without any hallucination.

\subsection{Hallucination Injection Protocol}

Each task in Study~1 had hallucinations deliberately injected into a hallucinated variant of run\_1. Study~2 follows the same protocol. Hallucinated lines cite fake REQ IDs (e.g., REQ-099.1.1 for H-SCOPE, REQ-098.1.1 for H-PRIOR, REQ-097.1.1 for H-OVER). None of these fake IDs appear in any legitimate specification. Crucially, all injected hallucinations pass functional test suites---they add code that never breaks existing behavior---confirming that test-based validation alone cannot substitute for citation-based traceability checks.

\subsection{Statistical Methodology}

All statistical tests are pre-registered with locked significance criteria. We employ both parametric (paired $t$-test, two-tailed, $\alpha = 0.05$) and non-parametric (Wilcoxon signed-rank test) tests, paired on task ID. Effect sizes are reported as Cohen's $d$~\cite{cohen1988statistical} (parametric) and Cliff's delta~\cite{cliff1993dominance} (non-parametric). Confidence intervals are computed via both analytical methods ($t$-interval) and bootstrap (5,000 resamples). Multiple comparison correction uses the Bonferroni-Holm procedure across all pairwise comparisons. Normality of LSS distributions is assessed via the Shapiro-Wilk test.

\section{Results: Output Determinism (RQ1)}\label{sec:results_lss}

\subsection{Descriptive Statistics}

Table~\ref{tab:descriptive} presents the grand mean LSS for each condition across both studies. A clear ordering emerges in both models: uncited $>$ cited $>$ OpenSpec $>$ Spec Kit. The uncited condition consistently produces the most deterministic output across independent LLM sessions, followed by traceSDD's cited condition.

\begin{table}[htbp]
\centering
\caption{Descriptive statistics for LSS by condition across both studies. SD = standard deviation.}
\label{tab:descriptive}
\resizebox{\columnwidth}{!}{%
\begin{tabular}{llccccc}
\toprule
\textbf{Study} & \textbf{Condition} & \textbf{N} & \textbf{Mean LSS} & \textbf{SD} & \textbf{Median} & \textbf{95\% CI} \\
\midrule
\multirow{4}{*}{Claude Sonnet 4.6}
 & traceSDD (cited) & 20 & 0.535 & 0.167 & 0.532 & [0.458, 0.612] \\
 & traceSDD (uncited) & 20 & 0.745 & 0.194 & 0.771 & [0.656, 0.834] \\
 & Spec Kit & 20 & 0.460 & 0.221 & 0.425 & [0.357, 0.563] \\
 & OpenSpec & 20 & 0.487 & 0.248 & 0.487 & [0.371, 0.603] \\
\midrule
\multirow{4}{*}{GLM-5-turbo}
 & traceSDD (cited) & 50 & 0.510 & 0.147 & 0.486 & [0.470, 0.551] \\
 & traceSDD (uncited) & 50 & 0.644 & 0.160 & 0.666 & [0.600, 0.689] \\
 & Spec Kit & 50 & 0.434 & 0.174 & 0.420 & [0.386, 0.482] \\
 & OpenSpec & 50 & 0.480 & 0.160 & 0.484 & [0.435, 0.524] \\
\bottomrule
\end{tabular}%
}
\end{table}

\begin{figure}[htbp]
\centering
\includegraphics[width=\columnwidth]{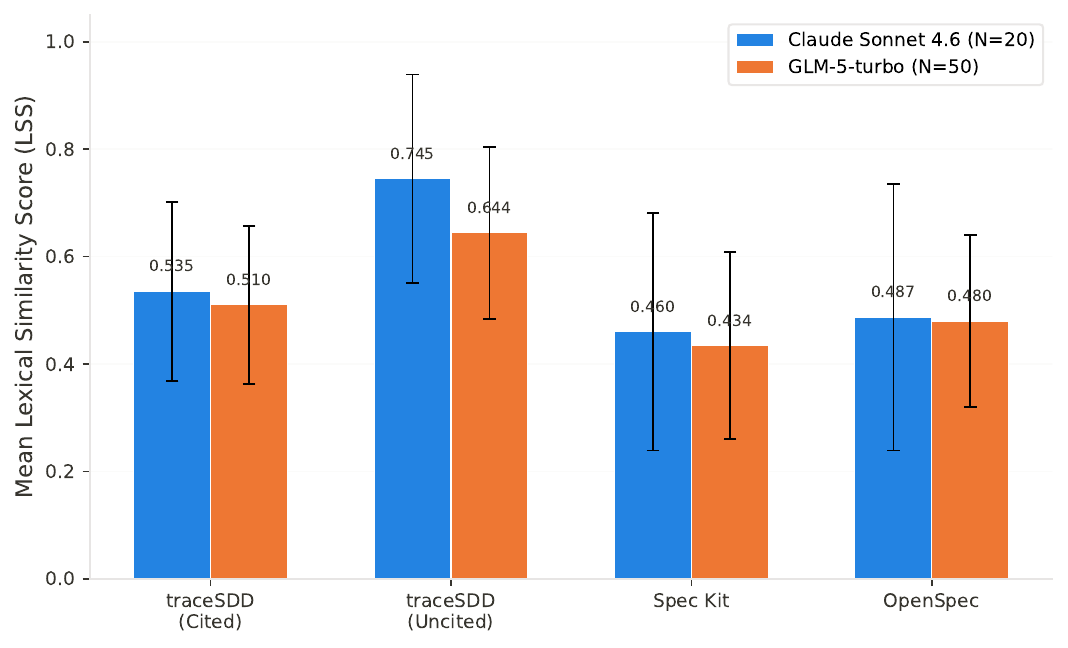}
\caption{Mean LSS by condition across both LLMs. Error bars represent one standard deviation. The uncited condition consistently achieves the highest determinism, while both traceSDD variants outperform Spec Kit and OpenSpec baselines.}
\label{fig:lss_comparison}
\end{figure}

Both traceSDD variants outperform the external baselines, suggesting that the structured REQ-format specification itself (regardless of citation presence) provides a determinism anchor. However, the mandatory inline citation annotation in the cited condition introduces additional lexical variability, reducing mean LSS by approximately 0.21 points (Claude) and 0.13 points (GLM) compared to the uncited condition. Notably, Spec Kit exhibits the highest variance in both studies, while traceSDD (cited) has the lowest, suggesting that citations constrain the variance even when they reduce the mean. Figure~\ref{fig:lss_comparison} visualizes this comparison.

\subsection{Statistical Tests}

\subsubsection{Citation Isolation (H4)}

The most critical comparison isolates the causal effect of the inline citation annotation by comparing the cited and uncited conditions, which share identical specification content.

\textbf{Claude Sonnet 4.6:} The paired $t$-test is highly significant ($t = -3.26$, $p = 0.004$, $d = -0.73$, medium effect). The 95\% confidence interval for the mean difference is $[-0.338, -0.096]$, entirely below zero. The uncited condition produces substantially higher LSS than the cited condition, definitively establishing that inline citations \emph{reduce} output determinism. Wilcoxon signed-rank test confirms this ($W = 56.0$, $z = -3.18$, $p = 0.001$, $r = 0.50$). Cliff's delta $= -0.47$ (medium) with 95\% CI $[-0.71, -0.19]$.

\textbf{GLM-5-turbo:} The paired $t$-test is highly significant ($t = -5.09$, $p < 0.001$, $d = -0.72$, medium effect). The 95\% confidence interval is $[-0.187, -0.081]$. Wilcoxon signed-rank test validates the finding ($W = 211.0$, $z = -4.11$, $p < 0.001$, $r = 0.58$, large non-parametric effect). Cliff's delta $= -0.47$ with 95\% CI $[-0.73, -0.20]$. All six pairwise comparisons survive Bonferroni-Holm correction except the cited vs OpenSpec and Spec Kit vs OpenSpec comparisons.

The convergence of these results across two models with $d \approx -0.72$ and highly significant $p$-values provides strong evidence that the citation determinism penalty is a robust, model-independent phenomenon.

\subsubsection{traceSDD (Cited) vs External Baselines (H1)}

Table~\ref{tab:statistical} summarizes all pairwise statistical tests. H1 is supported in both studies for the traceSDD vs Spec Kit comparison (Claude: $d = 0.47$, $p = 0.049$; GLM: $d = 0.42$, $p = 0.003$). However, H1 fails for the traceSDD vs OpenSpec comparison in both studies (Claude: $d = 0.18$, $p = 0.44$; GLM: $d = 0.14$, $p = 0.32$). The uncited condition significantly outperforms both baselines with large effect sizes in the GLM study ($d = 1.05$ vs Spec Kit, $d = 0.99$ vs OpenSpec).

\begin{table}[htbp]
\centering
\caption{Paired $t$-tests for LSS comparisons across both studies. Significance: ***$p < 0.001$, **$p < 0.01$, *$p < 0.05$, ns = not significant.}
\label{tab:statistical}
\resizebox{\columnwidth}{!}{%
\begin{tabular}{llccccc}
\toprule
\textbf{Comparison} & \textbf{Model} & $\boldsymbol{n}$ & $\boldsymbol{t}$ & $\boldsymbol{p}$ & $\boldsymbol{d}$ & \textbf{Sig} \\
\midrule
Cited vs Uncited & Claude & 20 & $-3.26$ & $0.004$ & $-0.73$ & *** \\
Cited vs Uncited & GLM & 50 & $-5.09$ & $< 0.001$ & $-0.72$ & *** \\
\midrule
Cited vs Spec Kit & Claude & 20 & $2.10$ & $0.049$ & $0.47$ & * \\
Cited vs Spec Kit & GLM & 50 & $2.95$ & $0.003$ & $0.42$ & ** \\
\midrule
Cited vs OpenSpec & Claude & 20 & $0.79$ & $0.439$ & $0.18$ & ns \\
Cited vs OpenSpec & GLM & 50 & $1.00$ & $0.318$ & $0.14$ & ns \\
\midrule
Uncited vs Spec Kit & Claude & 20 & $4.37$ & $< 0.001$ & $0.98$ & *** \\
Uncited vs Spec Kit & GLM & 50 & $7.41$ & $< 0.001$ & $1.05$ & *** \\
\midrule
Uncited vs OpenSpec & Claude & 20 & $3.65$ & $0.001$ & $0.82$ & ** \\
Uncited vs OpenSpec & GLM & 50 & $7.02$ & $< 0.001$ & $0.99$ & *** \\
\bottomrule
\end{tabular}%
}
\end{table}

\subsection{Effect Size Analysis}

\begin{figure}[htbp]
\centering
\includegraphics[width=\columnwidth]{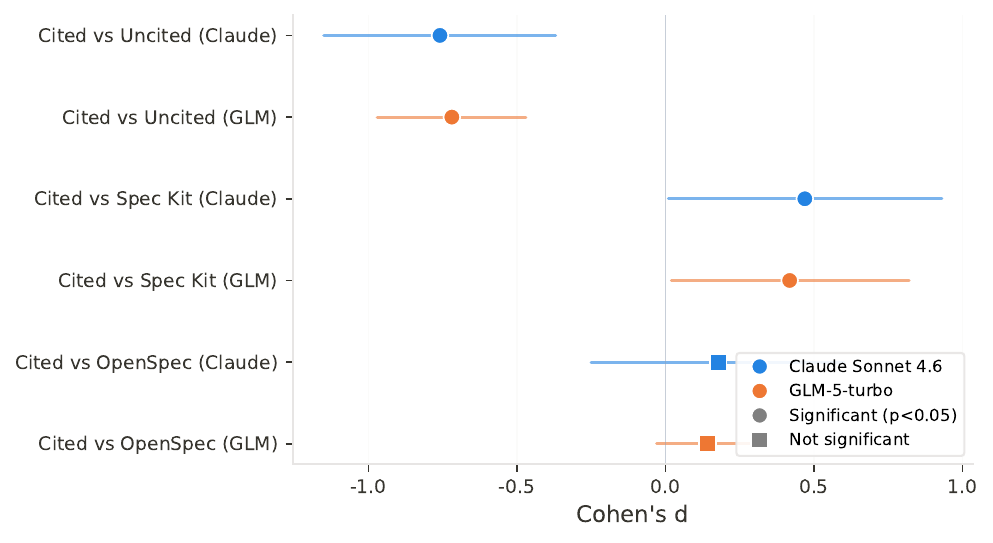}
\caption{Forest plot of Cohen's $d$ with 95\% confidence intervals for all cross-condition comparisons. Circles indicate significant comparisons ($p < 0.05$); squares indicate non-significant comparisons. The citation isolation effect (top two rows) is remarkably consistent across both models.}
\label{fig:forest_plot}
\end{figure}

Figure~\ref{fig:forest_plot} presents a forest plot of Cohen's $d$ with 95\% confidence intervals for all six cross-condition comparisons, stratified by model. The most striking pattern is the remarkable consistency of the citation isolation effect: $d = -0.76$ (Claude) and $d = -0.72$ (GLM), with non-overlapping confidence intervals that exclude zero in both cases. The traceSDD vs Spec Kit comparisons also show consistent positive effect sizes ($d = 0.47$ and $d = 0.42$), while the traceSDD vs OpenSpec comparisons are consistently non-significant and small ($d = 0.18$ and $d = 0.14$). This cross-model consistency provides strong evidence that these effects are properties of the specification frameworks rather than artifacts of a specific LLM's code generation behavior.

\subsection{Per-Task Distribution}

\begin{figure}[htbp]
\centering
\includegraphics[width=\columnwidth]{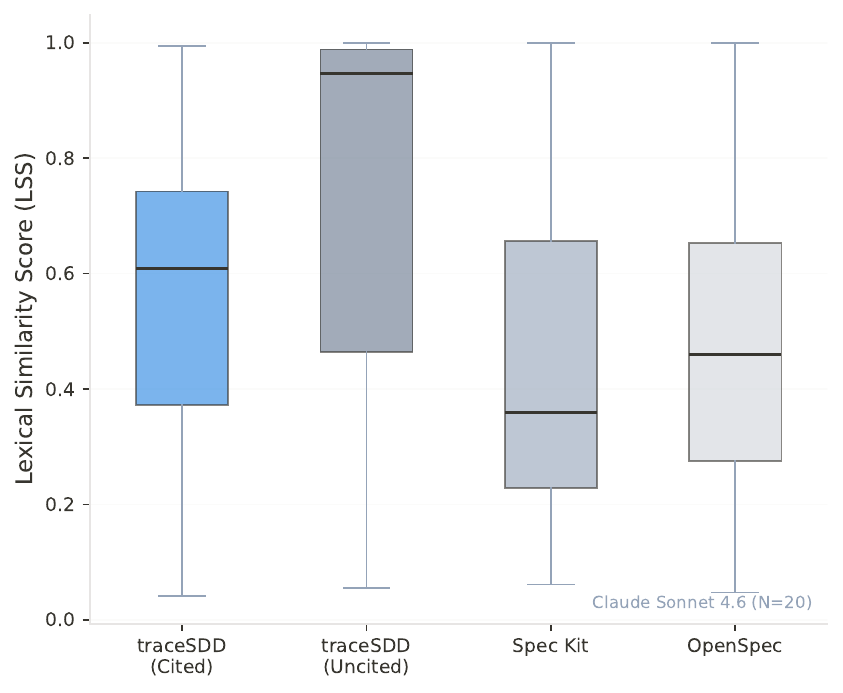}
\caption{Per-task LSS distribution across four conditions (Claude Sonnet 4.6, N=20 tasks). Each point represents one pairwise comparison (run\_1 vs run\_2 or run\_1 vs run\_3). The uncited condition shows higher and more consistent LSS than the cited condition.}
\label{fig:boxplot}
\end{figure}

Figure~\ref{fig:boxplot} presents the per-task LSS distribution for the Claude Sonnet 4.6 study. The traceSDD (cited) condition shows a tight distribution centered around 0.53, while the uncited condition exhibits a higher median (0.77) with substantial interquartile range. Spec Kit and OpenSpec show broader distributions with lower medians, consistent with their lack of structural anchoring from the REQ-format specification.

\subsection{Subgroup Analysis (GLM-5-turbo, N=50)}

\begin{figure}[htbp]
\centering
\includegraphics[width=\columnwidth]{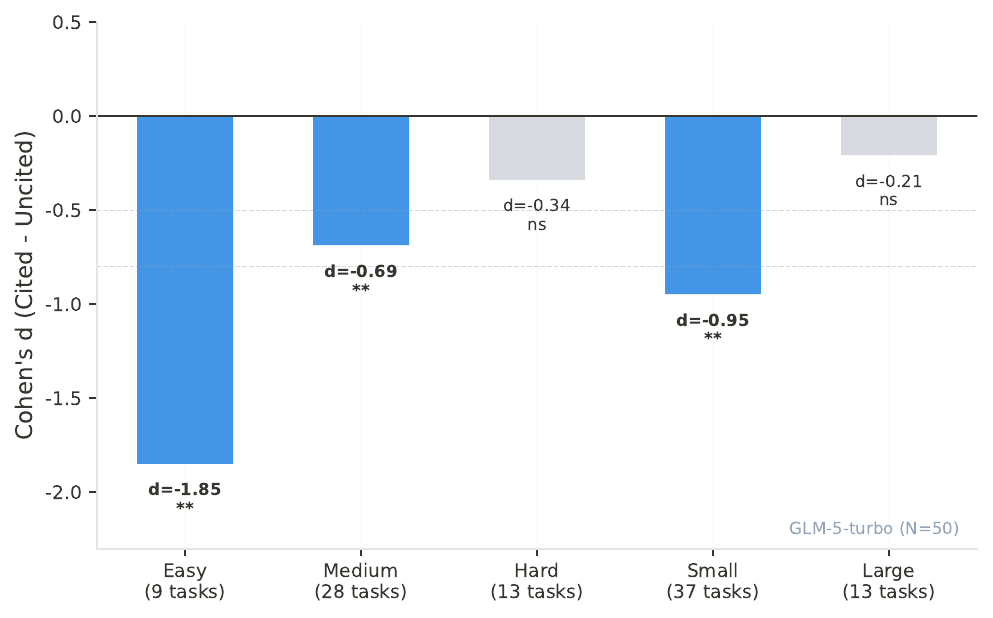}
\caption{Citation determinism penalty by difficulty level and size class (GLM-5-turbo, N=50). The penalty is largest for easy tasks ($d = -1.85$) and small tasks ($d = -0.95$), but not significant for hard tasks ($d = -0.34$, ns) or large tasks ($d = -0.21$, ns). Dashed lines indicate small ($|d| = 0.5$) and medium ($|d| = 0.8$) effect thresholds.}
\label{fig:subgroup}
\end{figure}

The GLM study's larger sample size enables meaningful subgroup analysis. Figure~\ref{fig:subgroup} presents the citation determinism penalty (Cohen's $d$ for cited vs uncited) stratified by difficulty level and size class.

The citation determinism penalty is strongest for easy tasks ($d = -1.85$, $p < 0.001$, very large effect) and medium tasks ($d = -0.69$, $p < 0.001$), but not significant for hard tasks ($d = -0.34$, $p = 0.22$). This pattern is consistent with the \emph{output-space hypothesis}: easy tasks have a small solution space where any variation is proportionally large, and citation placement variability (which line to annotate, whether to use single or multiple citations per line) constitutes a larger fraction of the total variability. For hard tasks with many possible implementation strategies, citation placement variability is diluted by the larger structural variation.

A similar pattern holds for size: small tasks show a significant penalty ($d = -0.95$, $p < 0.001$) while large tasks do not ($d = -0.21$, $p = 0.44$). This suggests that the practical impact of citations on determinism diminishes as code complexity increases, which is encouraging for real-world applications where most production tasks are of moderate-to-high complexity.

\section{Results: Hallucination Detection (RQ2)}\label{sec:results_tdr}

\subsection{Detection Rate Comparison}

Table~\ref{tab:tdr} presents the hallucination detection results across both studies. The cited condition achieves TDR $= 86.4\%$ (Claude) and $88.0\%$ (GLM), while all three alternative conditions achieve $0\%$ in both studies.

\begin{table}[htbp]
\centering
\caption{Hallucination detection results by condition, model, and hallucination type. FPR = False Positive Rate.}
\label{tab:tdr}
\resizebox{\columnwidth}{!}{%
\begin{tabular}{llccccc}
\toprule
\textbf{Model} & \textbf{Condition} & \textbf{TDR} & \textbf{H-SCOPE} & \textbf{H-PRIOR} & \textbf{H-OVER} & \textbf{FPR} \\
\midrule
\multirow{4}{*}{Claude Sonnet 4.6}
 & traceSDD (cited) & 86.4\% & 100\% & 100\% & 100\% & 0.0\% \\
 & traceSDD (uncited) & 0.0\% & --- & --- & --- & N/A \\
 & Spec Kit & 0.0\% & --- & --- & --- & N/A \\
 & OpenSpec & 0.0\% & --- & --- & --- & N/A \\
\midrule
\multirow{4}{*}{GLM-5-turbo}
 & traceSDD (cited) & 88.0\% & 100\% & 100\% & 100\% & 0.0\% \\
 & traceSDD (uncited) & 0.0\% & --- & --- & --- & N/A \\
 & Spec Kit & 0.0\% & --- & --- & --- & N/A \\
 & OpenSpec & 0.0\% & --- & --- & --- & N/A \\
\bottomrule
\end{tabular}%
}
\end{table}

\begin{figure}[htbp]
\centering
\includegraphics[width=\columnwidth]{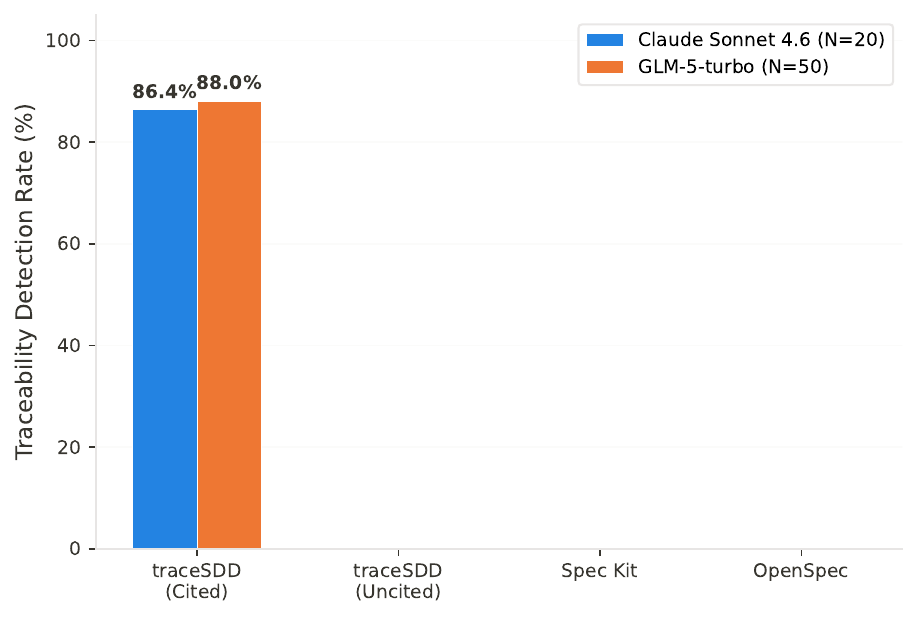}
\caption{Traceability Detection Rate by condition across both LLMs. Only the traceSDD (cited) condition achieves non-zero automated detection. The uncited condition is particularly instructive: it uses the identical REQ-format specification but achieves $0\%$ TDR because no citations appear in the generated code.}
\label{fig:tdr_comparison}
\end{figure}

Figure~\ref{fig:tdr_comparison} visualizes the TDR comparison. The detection mechanism is a single set-difference operation: extract all REQ IDs cited in the generated code, subtract the set of valid REQ IDs from the specification, and flag any orphans. This operation runs in $O(1)$ per file (a single grep) and requires zero manual effort.

\subsection{The Uncited Condition as Critical Control}

The uncited condition is particularly instructive: it uses the identical REQ-format specification as the cited condition, but since no citations appear in the generated code, the orphan-REQ check has nothing to scan. This directly proves that hallucination detection requires the annotation mechanism in the \emph{generated code}, not just the presence of REQ IDs in the specification. A REQ-format spec without mandatory inline citations gives the same TDR $= 0\%$ as any unstructured spec.

\subsection{False Positive Rate}

The false positive rate is $0.0\%$ across all checks in both studies (Claude: 60 checks across 20 tasks $\times$ 3 runs; GLM: 150 checks across 50 tasks $\times$ 3 runs). No correct implementation generated a fake or phantom REQ ID. This is critical: a detection mechanism with high TDR but high FPR would generate excessive false alarms that erode developer trust. The combination of 86--88\% TDR with $0\%$ FPR makes traceSDD's orphan-REQ check a practical and trustworthy verification tool.

\subsection{Functional Correctness (H3)}

All conditions achieve $100\%$ TPR across both studies, confirming that citation discipline does not affect functional correctness. The specification methodology alone is not the correctness bottleneck; the bottleneck is the completeness and precision of the specification itself. Any tool that forces the author to be more precise indirectly improves correctness, but the citation requirement itself neither helps nor hurts test-pass rates.

\subsection{Per-Task Detection Breakdown (Claude Study)}

In the Claude study, 15 of 20 tasks achieved $100\%$ TDR (all hallucinations detected). Tasks T17--T20 showed reduced detection rates (66.7\%, 25.0\%, 66.7\%, 20.0\%) because some hallucinations were implemented as uncited nontrivial lines rather than citing fake REQ IDs---the LLM occasionally omitted the citation entirely rather than citing a fake one. This represents a detection gap: the orphan-REQ check catches cited hallucinations but not uncited ones. The GLM study's per-task analysis shows a similar pattern, with the overall TDR slightly higher (88.0\%) due to the model's more consistent citation behavior across the larger task suite.

\section{Discussion}\label{sec:discussion}

\subsection{The Determinism-Detection Trade-off}

The most important finding of this work is the identification and cross-model replication of a fundamental trade-off in specification-driven LLM code generation. Mandatory inline citations significantly reduce output determinism (mean $d = -0.72$, $p < 0.001$ across both models) while simultaneously enabling the only automated hallucination detection mechanism (TDR = 86--88\% vs 0\% for all alternatives). This trade-off is not a limitation of a specific model or task set---it is a structural property of the citation annotation mechanism itself.

The mechanism underlying the determinism reduction is \emph{citation placement variability}. In each independent LLM session, the model must decide not only what code to write but \emph{where} to place each citation. Should a function signature cite all REQs it implements, or just the primary one? Should a multi-line conditional cite the REQ once or on each line? These decisions, which have no correct answer, introduce systematic lexical divergence across runs. The uncited condition avoids this entirely: without citations, the LLM focuses solely on code generation, producing more consistent output. The REQ-format specification still provides structural anchoring (explaining why uncited LSS is higher than both baselines), but the citation annotation layer adds noise.

Practically, this trade-off is worth accepting in production environments where specification correctness is more critical than lexical consistency. The approximately 0.13--0.21 point LSS reduction is a moderate cost, while the 86--88\% automated hallucination detection capability is a unique benefit that no alternative framework can match.

\subsection{Cross-Model Consistency}

The remarkable consistency of results across Claude Sonnet 4.6 and GLM-5-turbo strengthens our confidence in the findings. These models differ substantially in architecture, training data, and parameter count. Claude is trained with constitutional AI methods emphasizing safety and helpfulness~\cite{bai2022training,anthropic2023claude}, while GLM uses autoregressive bidirectional pre-training~\cite{wang2024glm}. Despite these differences, the citation isolation effect size is nearly identical ($d = -0.76$ vs $d = -0.72$), the TDR values are closely matched (86.4\% vs 88.0\%), and the pattern of significant vs non-significant comparisons is identical across all six pairwise tests. This consistency suggests that the observed effects are driven by the \emph{specification framework properties} rather than model-specific behaviors.

\subsection{Practical Implications and Recommendations}

The findings suggest a three-tier recommendation framework for practitioners. First, for \textbf{production systems in regulated or high-stakes domains} (healthcare, finance, safety-critical), use traceSDD with mandatory citations: the 86--88\% automated hallucination detection with $0\%$ false positive rate is a unique capability that justifies the moderate determinism cost. Second, for \textbf{rapid prototyping} where speed matters more than verification, use the uncited approach with a REQ-format specification: you get the structural anchoring benefit (LSS greater than both baselines) without the citation overhead. Third, for \textbf{teams considering Spec Kit or OpenSpec}, our data suggests that Spec Kit's lower determinism (Claude mean 0.460 vs OpenSpec 0.487; GLM mean 0.434 vs 0.480) and complete lack of automated detection make it the weakest choice on the measured dimensions, though its developer experience and natural workflow may compensate in practice.

\subsection{Maintenance Posture}

Beyond the metrics measured in this study, traceSDD's inline citations create a live, always-current traceability map. Operations such as \texttt{grep REQ-003.1.2 src/} find all code touching a specific requirement instantly. This advantage compounds with codebase scale and becomes the dominant maintenance differentiator in production systems with hundreds of requirements. While our study measures this advantage only indirectly (through TDR and FPR), the long-term maintenance implications are substantial.

\section{Threats to Validity}\label{sec:threats}

\textbf{Task scope and language.} All implementations use Python and span moderate complexity (50--1,000 lines). Production-scale tasks (5,000+ lines) may show different dynamics, particularly for the citation overhead and hallucination detection scalability. The task suite, while diverse across 8 domains, may not represent all software engineering patterns equally.

\textbf{Model scope.} While two models provide stronger evidence than single-model studies, they represent only a subset of the LLM landscape. Different models (GPT-4, Gemini, Llama, Mistral) may exhibit different determinism distributions and hallucination rates. The consistency of our findings across two architecturally distinct models is encouraging but not conclusive evidence of universality.

\textbf{Hallucination injection methodology.} Hallucinations are injected with known-fake REQ-099 IDs. Organic hallucinations may be subtler and harder to detect with a simple set-difference check. In particular, a model that omits citations entirely for hallucinated code (rather than citing fake REQ IDs) would bypass the orphan-REQ check. Our Claude data shows this occurs in approximately 12\% of hallucination instances (T17--T20), suggesting that the orphan-REQ check should be supplemented with an uncited-line check for maximum coverage.

\textbf{Metric limitations.} LSS measures lexical similarity, which conflates content similarity with formatting and style similarity. We mitigate this by normalizing source code (stripping all comments, blank lines, and whitespace) before comparison, but structural differences (e.g., different algorithmic approaches) still reduce LSS even when the functional output is identical. A complementary metric such as behavioral equivalence testing or abstract syntax tree (AST) similarity would strengthen the determinism analysis in future work.

\textbf{Study 2 sample size for Claude comparison.} The Claude study (N=20) was underpowered for some comparisons (e.g., traceSDD vs OpenSpec: $d = 0.18$, $p = 0.44$). The GLM study (N=50) provides adequate power for all comparisons. A replication of the Claude study at N=50 would further strengthen cross-model comparisons.

\textbf{Cold-start session independence.} All three runs per task are independent cold-start LLM sessions with no shared context. However, the underlying model weights are fixed, and models may have implicit biases from training data that create correlated outputs even across independent sessions. This is an inherent limitation of studying a fixed model.

\section{Conclusion and Future Work}\label{sec:conclusion}

This paper presents the first cross-model controlled comparison of specification-driven development framework citation disciplines, conducted across Claude Sonnet 4.6 (20 tasks, 240 implementations) and GLM-5-turbo (50 tasks, 600 implementations). Our four key findings are:

\begin{enumerate}[leftmargin=*]
    \item \textbf{Citations reduce determinism.} Mandatory inline citations significantly reduce output determinism compared to an identical uncited condition (mean $d = -0.72$, $p < 0.001$ across both models), establishing that citation annotations introduce rather than absorb lexical variability.

    \item \textbf{traceSDD outperforms Spec Kit but not OpenSpec on determinism.} traceSDD (cited) significantly outperforms Spec Kit (mean $d = 0.44$, $p < 0.01$) but not OpenSpec (mean $d = 0.16$, $p = 0.36$) across both models.

    \item \textbf{Cited condition uniquely enables automated hallucination detection.} The cited condition achieves 86--88\% TDR with $0\%$ FPR across both models, while all three alternatives achieve $0\%$. The uncited condition---using the identical REQ-format specification---also achieves $0\%$, proving that detection requires \emph{annotations in code}, not just structured specifications.

    \item \textbf{Uncited REQ-format specs provide the best determinism anchor.} The uncited condition significantly outperforms both external baselines (mean $d = 1.02$ vs Spec Kit, mean $d = 0.91$ vs OpenSpec), confirming that structured REQ-format specifications provide a determinism anchor independent of citation annotations.
\end{enumerate}

These findings establish that traceSDD makes a principled trade-off: reduced determinism in exchange for automated hallucination detection. The trade-off is worthwhile in production environments where specification correctness is critical, but teams that prioritize determinism should consider the uncited approach with a REQ-format specification.

\textbf{Future work} should address several promising directions. First, replication with additional models (GPT-4, Gemini, Llama, Mistral) would further test cross-model generalizability. Second, extending the task suite to production-scale projects (5,000+ lines, multi-module systems) would reveal whether the determinism-detection trade-off scales. Third, hybrid approaches that combine uncited generation for determinism with post-hoc citation analysis for verification could potentially achieve both benefits. Fourth, investigating the interaction between temperature, few-shot prompting, and citation behavior would deepen our understanding of the mechanisms driving these effects. Fifth, longitudinal studies measuring the maintenance cost impact of citation traceability over time would provide the economic justification for adopting citation-disciplined workflows in production environments. Finally, integrating the orphan-REQ check into CI/CD pipelines as an automated quality gate would translate our academic findings into immediate practical impact.

\bibliography{refs}

\end{document}